\documentclass{article}
\makeatletter
\def\blfootnote{\xdef\@thefnmark{}\@footnotetext}
\makeatother
\usepackage{amsmath,amssymb,amsfonts}
\begin{document}
\newtheorem{theorem}{Theorem}[section]
\newtheorem{lemma}[theorem]{Lemma}
\newtheorem{conjecture}[theorem]{Conjecture}
\newtheorem{corollary}[theorem]{Corollary}
\newtheorem{definition}[theorem]{Definition}
\newtheorem{assumption}[theorem]{Assumption}
\newtheorem{proposition}[theorem]{Proposition}
\newtheorem{problem}[theorem]{Problem}
\newtheorem{remark}[theorem]{Remark}
\newtheorem{example}[theorem]{Example}
\def\emptyset{\varnothing}
\def\setminus{\smallsetminus}
\def\Diff{{\mathrm{Diff}}}
\def\Vir{{\mathrm{Vir}}}
\def\Aut{{\mathrm{Aut}}}
\def\End{{\mathrm{End}}}
\def\Ker{{\mathrm{Ker}}}
\def\Mor{{\mathrm{Mor}}}
\def\opp{{\mathrm{opp}}}
\def\Rep{{\mathrm{Rep}}}
\def\Tr{{\mathrm{Tr}}}
\def\tr{{\mathrm{tr}}}
\def\supp{{\mathrm{supp}}}
\def\ind{{\mathrm{ind}}}
\def\gen{{\mathrm{gen}}}
\def\id{{\mathrm{id}}}
\def\dual{{\mathrm{dual}}}
\def\ch{{\mathrm{ch}}}
\def\A{{\mathcal{A}}}
\def\B{{\mathcal{B}}}
\def\C{{\mathbb{C}}}
\def\E{{\mathcal{E}}}
\def\G{{\mathcal{G}}}
\def\K{{\mathcal{K}}}
\def\LL{{\mathcal{L}}}
\def\M{{\mathcal{M}}}
\def\N{{\mathbb{N}}}
\def\OO{{\mathcal{O}}}
\def\R{{\mathbb{R}}}
\def\U{{\mathcal{U}}}
\def\Q{{\mathbb{Q}}}
\def\Z{{\mathbb{Z}}}
\def\a{{\alpha}}
\def\e{{\varepsilon}}
\def\g{{\mathfrak g}}
\def\la{{\lambda}}
\def\si{{\sigma}}
\def\de{{\delta}}
\def\isom{{\cong}}
\newcommand{\Ad}{\mathop{\mathrm{Ad}}\nolimits}
\newcommand{\Hom}{\mathop{\mathrm{Hom}}\nolimits}
\newcommand{\coker}{\mathop{\mathrm{coker}}\nolimits}
\def\qed{{\unskip\nobreak\hfil\penalty50
\hskip2em\hbox{}\nobreak\hfil$\square$
\parfillskip=0pt \finalhyphendemerits=0\par}\medskip}
\def\proof{\trivlist \item[\hskip \labelsep{\bf Proof.\ }]}
\def\endproof{\null\hfill\qed\endtrivlist\noindent}

\title{Conformal Field Theory,\\
Vertex Operator Algebras and\\
Operator Algebras}
\author{
{\sc Yasuyuki Kawahigashi}\\
{\small Graduate School of Mathematical Sciences}\\
{\small The University of Tokyo, Komaba, Tokyo, 153-8914, Japan}
\\[0,05cm]
{\small and}
\\[0,05cm]
{\small Kavli IPMU (WPI), the University of Tokyo}\\
{\small 5-1-5 Kashiwanoha, Kashiwa, 277-8583, Japan}\\
{\small e-mail: {\tt yasuyuki@ms.u-tokyo.ac.jp}}}
\maketitle{}
\begin{abstract}
We present recent progress in theory of local
conformal nets which is
an operator algebraic approach to study chiral
conformal field theory.  We emphasize representation
theoretic aspects and relations to theory of
vertex operator algebras which gives a different
and algebraic formulation of chiral conformal
field theory.
\end{abstract}

\section{Introduction}

Quantum field theory is a vast area in physics and
two-dimensional conformal field theory 
has caught much attention recently.  A two-dimensional
conformal field theory decomposes into two chiral
conformal field theories, and here we present
mathematical studies of a chiral conformal field theory
based on operator algebras.  It is within a scope of
what is called algebraic quantum field theory and
our mathematical object is called a local conformal net.

The key idea in algebraic quantum field theory is to work
on operator algebras generated by observables in a
spacetime region rather than quantum fields.  In chiral
conformal field theory, the spacetime becomes a one-dimensional
circle and a spacetime region is an interval in it, which
is a nonempty, nondense, open and connected set in the
circle, so we deal with a continuous family of operator
algebras parameterized by intervals.  This is what a
local conformal net is.

Each operator algebra of a local conformal net acts
on the same Hilbert space from the beginning, but we also
consider its representation theory on another Hilbert
space.  Such a representation corresponds to a notion
of a charge, and a unitary equivalence class of a
representation is called a superselection sector.
In the Doplicher-Haag-Roberts theory \cite{DHR},
a representation is realized with a DHR endomorphism
of one operator algebra, and such an endomorphism
produces a subfactor in the sense of the Jones theory
\cite{J1}, \cite{J2}.  Subfactor theory plays
an important role in this approach.  It has revolutionized
theory of operator algebras and revealed its surprising
deep relations to 3-dimensional topology, quantum groups and
solvable lattice models.  Its connection to quantum field theory
was clarified by Longo \cite{L1} and it has been
an important tool also in conformal field theory since then.

Representation theory of a local conformal net gives a
powerful tool to study chiral conformal field theory.
We present $\alpha$-induction, a certain induction
procedure for representation theory of a local conformal
net, and its use for classification theory.

A vertex operator algebra gives another axiomatization of
a chiral conformal field theory and it has started with
the famous Moonshine conjecture \cite{CN}.  The axiomatic
framework has been established in \cite{FLM} and we have
had many research papers on this topic.  This is an
algebraic axiomatization of Fourier expansions of
a family of operator-valued distributions on the
one-dimensional circle.  Since a local conformal net
and a vertex operator algebra give different axiomatizations
of the same physical theory, it is natural to expect that
they have many common features.  There have been many
parallel results in the two theories, but a precise
relation between the two were not known until recently.
We have established that if a vertex operator algebra
satisfies unitarity and an extra mild assumption called
strong locality, then we can construct the corresponding
local conformal net and also recover the original vertex
operator algebra from the local conformal net.  Strong
locality is known to be satisfied for most examples
and we do not know any example of a vertex operator
algebra which does not have strong locality.

There are many open problems to study in the operator
algebraic approach to chiral conformal field theory.
We present some of them in this article.

We refer a reader to lecture notes \cite{K1} for more
details with an extensive bibliography.

This work was supported in part by 
Research Grants and the Grants-in-Aid
for Scientific Research, JSPS.

\section{Algebraic quantum field theory and local
conformal nets}

In a common approach to quantum field theory such as the
Wightman axioms, we deal with quantum fields which are
a certain kind of operator-valued distributions on
the spacetime acting on the same Hilbert space
together with a spacetime symmetry group.
An operator-valued distribution $T$ applied to a test
function $f$ gives $\langle T, f\rangle$ which is an
(often unbounded) operator.  Handling distributions,
rather than functions, and unbounded operators causes
technical difficulties, so an idea of algebraic
quantum field theory of Haag-Kastler
is to study operator algebras
generated by observables in a spacetime region.
Let $T$ be an operator-valued distribution and $f$
be a test function supported in $O$ which is
a spacetime region.  Then $\langle T,f\rangle$
gives an observable in $O$ (if it is self-adjoint).
Let $A(O)$ be the von Neumann algebra generated
by these observables.  
(A von Neumann algebra is an algebra of bounded linear
operators on a Hilbert space containing the identity
operator which is closed under the $*$-operation and
the strong operator topology.  This topology is given
by pointwise convergence on the Hilbert space.)
We have a family $\{A(O)\}$
of von Neumann algebras.  We impose physically natural
axioms on such a family and make a mathematical
study of these axioms.

We apply the above general idea to 2-dimensional
conformal field theory.  We first consider the
2-dimensional Minkowski space with space coordinate
$x$ and time coordinate $t$.
We have a certain restriction procedure of a conformal
field theory on the Minkowski space to the two light
rays $\{x=\pm t\}$.  In this way, we can regard one light
ray as a kind of spacetime though it has only one dimension.
Then conformal symmetry can move the point at infinity of
this light ray, so our space should be now $S^1$, the
one-point compactification of a light ray.  A spacetime
region is now a nonempty nondense open connected subset
of $S^1$ and such a set is called an interval.  Our 
mathematical object is a family of von Neumann algebras
$\{A(I)\}$ parametrized by an interval $I\subset S^1$.
We impose the following axioms on this family.

\begin{enumerate}
\item (Isotony)
For two intervals $I_1\subset I_2$, we have $A(I_1)\subset A(I_2)$.
\item (Locality) When two intervals $I_1, I_2$ are disjoint,
we have $[A(I_1), A(I_2)]=0$.
\item (M\"obius covariance)
We have a unitary representation $U$ of
$PSL(2, {\mathbb{R}})$ on the Hilbert space such that
we have $U(g) A(I) U(g)^*=  A(gI)$ for all
$g\in PSL(2,{\mathbb{R}})$, where $g$ acts on $S^1$
as a fractional linear transformation on $\R\cup\{\infty\}$ and 
$S^1\setminus\{-1\}$ is identified with $\R$ through
the Cayley transform $C(z)=-i(z-1)/(z+1)$.
\item (Conformal covariance) 
We have a projective unitary representation, still
denoted by $U$, of ${\mathrm{Diff}}(S^1)$, the group of
orientation preserving diffeomorphisms of $S^1$,
extending the unitary representation
$U$ of $PSL(2, \R)$ such that
\begin{align*}
U(g)\A(I) U(g)^* &= A(gI),\quad  g\in{\mathrm{Diff}}(S^1), \\
U(g)x U(g)^* &=x,\quad x\in A(I),\ g\in{\mathrm{Diff}}(I'),
\end{align*}
where $I'$ is the interior of the complement of $I$ and
${\mathrm{Diff}}(I')$ is the set of diffeomorphisms of $S^1$
which are the identity map on $I$.
\item (Positive energy condition)
The generator of the restriction of $U$ to the rotation
subgroup of $S^1$, the conformal Hamiltonian, is positive.
\item (Existence of the vacuum vector)
We have a unit vector $\Omega$, called the
vacuum vector, such that $\Omega$ is fixed
by the representation $U$ of 
$PSL(2, {\mathbb{R}})$ and
$(\bigvee_{I\subset S^1} A(I))\Omega$ is dense in the
Hilbert space,
where $\bigvee_{I\subset S^1} A(I)$ is the von Neumann
algebra generated by $A(I)$'s.
\item (Irreducibility) The von Neumann algebra
$\bigvee_{I\subset S^1}A(I)$  is the algebra of all
the bounded linear operators on the Hilbert space.
\end{enumerate}

Isotony is natural because a larger spacetime domain should
have more observables.  Locality comes from the Einstein
causality in the 2-dimensional Minkowski space that observables
in spacelike separated regions should commute with each other.
Note that we have a simple condition of disjointness instead
of spacelike separation.  Conformal covariance represents an
infinite dimensional symmetry.  This gives a reason for the name
``conformal'' field theory.  The positive energy condition
expresses positivity of the eigenvalues of the conformal
Hamiltonian.  The vacuum state is a physically distinguished
state of the Hilbert space.  Irreducibility means that our
Hilbert space is irreducible.

It is non-trivial to construct an example of a local conformal
net.  Basic sources of constructions are as follows.
These are also sources of constructions of vertex operator
algebras as we see below.

\begin{enumerate}
\item Affine Kac-Moody algebras \cite{FG}, \cite{W}, \cite{TL}
\item Virasoro algebra \cite{X2}, \cite{KL1}
\item Even lattices \cite{KL4}, \cite{DX}
\end{enumerate}

When we have some examples of local conformal nets, we have
the following methods to construct new ones.

\begin{enumerate}
\item Tensor product
\item Simple current extension \cite{BE}
\item Orbifold construction \cite{X3}
\item Coset construction \cite{X2}
\item Extension by a $Q$-system \cite{LR1}, \cite{KL1}, \cite{X4}
\end{enumerate}

The first four constructions were first studied for vertex
operator algebras.  The last one was first studied for
a local conformal net and later for vertex operator algebras
\cite{HKL}.  The Moonshine net, the operator algebraic
counterpart of the famous Moonshine vertex operator algebra,
is constructed from the Leech lattice, an even lattice of
rank 24, with a combination of the orbifold construction
and a simple current extension \cite{KL4}, for example.
This is actually given by a 2-step simple current extension
as in \cite{KS}.
The $Q$-system in the last construction was first
introduced in Longo \cite{L2}.  It is also known under the
name of a Frobenius algebra in algebraic literature.

Irreducibility implies
that each $A(I)$ has a trivial center.  Such an algebra
is called a factor.  It turns out that each algebra $A(I)$
is isomorphic to the Araki-Woods factor of type III$_1$
because the split property automatically holds by
\cite{MTW} and it implies hyperfiniteness of $A(I)$.
This shows that each single algebra
$A(I)$ contains no information about
a local conformal net and what is important is relative
relation among $A(I)$'s.

\section{Representation theory and superselection sectors}

We now present representation theory of a local conformal net.
Each algebra $A(I)$ of a local conformal net $\{A(I)\}$ acts
on the same Hilbert space having the vacuum vector from
the beginning, but we also consider a representation of
an algebra $A(I)$ on another common Hilbert space (without
a vacuum vector).

The Haag duality $A(I')=A(I)'$ automatically  holds
from the axioms, where the prime on the right hand
side denotes the commutant, and this implies that each
representation is represented with an endomorphism $\la$
of $A(I)$ for some fixed $I$.  This is a standard
Doplicher-Haag-Roberts theory adapted to a local
conformal net \cite{FRS}.    An endomorphism $\la$ produces
$\la(A(I))$ which is subalgebra of $A(I)$ and a factor, so it
is called a subfactor.
It is an object in the Jones theory of
subfactors \cite{J1}.  The relative size of the subfactor
$\la(A(I))$ with respect to $A(I)$ is called the  Jones
index $[A(I):\la(A(I))]$.  It turns out that the square
root $[A(I):\la(A(I))]^{1/2}$ of the Jones index gives a
proper notion of the dimension of the corresponding 
representation of $\{A(I)\}$ \cite{L1}.  The dimension
$\dim(\la)$ takes its value in the interval $[1,\infty]$.

It is important to have a notion of a tensor product
of representations of a local conformal net.
Note that while it is easy to define a tensor 
product of representations of a group, we have no
notion of a tensor product of representations of
an algebra.  It turns out that a composition of
endomorphisms of $A(I)$ for a fixed $I$ gives a
right notion of a tensor product of representations
\cite{DHR}.  In this way, we have a tensor category
of finite dimensional representations of $\{A(I)\}$.
The original action of $A(I)$ on the Hilbert space
is called the vacuum representation and has dimension 1.
It plays a role of a trivial representation.
In the original setting of the Doplicher-Haag-Roberts
theory on the higher dimensional Minkowski space, the
tensor product operation is commutative in a natural
sense and we have a symmetric tensor category.  Now
in the setting of chiral conformal field theory, the
commutativity is more subtle, and we have a 
structure of braiding \cite{FRS}.  We thus have a
braided tensor category of finite dimensional
representations.

We are often interested in a situation where we have
only finitely many irreducible representations and
such finiteness is usually called rational.  (This
rationality is well-studied in a context of
representation theory of quantum groups at roots
of unity in connection to quantum invariants in
3-dimensional topology.)  We have defined complete
rationality for a local conformal net, which means
we have only finitely many irreducible representations
up to unitary equivalence and all of them have finite
dimensions, and given its operator algebraic
characterization in terms of finiteness of the Jones index
of a certain subfactor in \cite{KLM}.  
(We originally assumed two more properties for complete
rationality, but they have been shown to be automatic
by \cite{LX}, \cite{MTW}, respectively.)
This characterization
is given by only studying the vacuum representation.
We have further proved that complete rationality implies
that the braiding of the representations is non-degenerate,
that is, we have the following theorem in \cite{KLM}.

\begin{theorem}
The tensor category of finite dimensional
representations of a completely rational
local conformal net is modular.
\end{theorem}

It is an important open problem to decide which modular
tensor category arises as the representation category
of a completely rational local conformal net.  The history
of classification theory of factors, group actions and
subfactors in theory of von Neumann algebras due to Connes,
Haagerup, Jones, Ocneanu and Popa culminating in \cite{P}
tells us
that as long as we have an analytic condition, generally
called amenability, we have no nontrivial obstruction to
realization of algebraic invariants.
This strongly suggests that any modular tensor category
is realized as the representation category of some
local conformal net, because we now have amenability automatically.
This conjecture has caught much 
attention these days because of recent work of Jones.
We turn to this problem again in the next section.

\section{Subfactors and tensor categories}

In the Jones theory of subfactors, we study an inclusion
$N\subset M$ of factors. In the original setting of Jones \cite{J1},
one considers type II$_1$ factors, but one has to deal with
type III factors in conformal field theory.  The Jones theory
has been extended to type III factors by Pimsner-Popa and Kosaki,
and many algebraic arguments are now more or less parallel in
the type II$_1$ and type III cases.  For simplicity, we assume
factors are of type II$_1$ in this section.
We refer reader to \cite{EK} for details of subfactor theory. 

We start with a subfactor $N\subset M$.  The Jones index $[M:N]$
is a  number in the interval $[1,\infty]$.  
In this section, we assume that the index is finite.
On the algebra $M$, we have the left and right actions of $M$
itself.  We restrict the left action to the subalgebra $N$,
and we have a bimodule ${}_N M_M$.  We make the completion
of $M$ with respect to the inner product arising from the
trace functional and obtain the Hilbert space $L^2(M)$.
For simplicity, we still write ${}_N M_M$ for this Hilbert
space with the left action of $N$ and the right action of $M$.
We make relative tensor powers such as
${}_N M\otimes_M M\otimes_N M \otimes_M\cdots$ and their 
irreducible decomposition gives four kinds of bimodules,
$N$-$N$, $N$-$M$, $M$-$N$ and $M$-$M$.
If we have only finitely many irreducible bimodules in this way,
we say that the subfactor $N\subset M$ is of finite depth.
In this case, finite direct sums of these irreducible $N$-$N$
(and $M$-$M$) bimodules (up to isomorphism)
give a fusion category.  Note that
the relative tensor product is not commutative in general
and we have no braiding structure.  

If we have a free action of
a finite group $G$ on a factor $M$, we have a subfactor
$N=M^G\subset M$.  The index is the order of $G$ and the 
fusion category of $N$-$N$ bimodules is the representation
category of $G$.  There are other constructions of
subfactors from actions of finite groups and their
quantum group versions give many interesting examples
of subfactors.  If the index is less than 4,
the set of all the possible values
is $\{4\cos^2 \pi/n\mid n=3,4,5,\dots\}$ \cite{J1}.
Classification of subfactors with index less than 4 has been
given in \cite{O1} and this is well-understood today 
in terms of quantum groups or conformal field theory.
Such classification of subfactors has been extended to
index value 5 \cite{JMS} recently.

There are some exceptional subfactors
which do not seem to arise from such constructions
involving (quantum) groups.  The most notable examples
are the Haagerup subfactor \cite{AH}, the Asaeda-Haagerup
subfactor \cite{AH} and the extended Haagerup subfactor
\cite{BMPS} in the index range $(4,5)$.
(The first two were constructed along an 
extension of the line of \cite{O1} and the last one is
based on the planar algebra of Jones.)
Such a subfactor produces an exceptional
fusion category and then it produces an exceptional
modular tensor category through the Drinfel$'$d center
construction.  (See \cite{I} for an operator algebraic
treatment of this.) Such a modular tensor category does not seem
to arise from a combination of other known constructions
applied to the Wess-Zumino-Witten models.
The above three subfactors were found through a combinatorial
search for a very narrow range of index values.  This strongly
suggests that there is a huge variety of exceptional fusion
categories and modular tensor categories beyond what is known
today.  History of classification theory of subfactors even
strongly suggests that there is a huge variety of exceptional
modular tensor categories even up to Witt equivalence ignoring
Drinfel$'$d centers, because it seems impossible to exhaust
all examples by prescribing construction methods.

As explained in the previous section, we strongly believe that
all such exceptional modular tensor categories do arise from
local conformal nets.  This would mean that there is a huge
variety of chiral conformal field theories beyond what is
known today.  For the Haagerup subfactor, a partial evidence
for this conjecture is given in \cite{EG}.

\section{$\a$-induction, modular invariants and
classification theory}

We next present an important tool to study representation of
a local conformal net.  For a subgroup $H$ of another group $G$
and a representation of $H$, we have a notion of an induced
representation of $G$.  We have some similar notion for a 
representation of a local conformal net.
Let $\{A(I)\subset B(I)\}$ be an inclusion of local conformal
nets and assume the index $[B(I):A(I)]$ is finite.  For a
a representation of $\{A(I)\}$ which is given by an endomorphism
$\la$ of a factor $A(I)$ for some fixed interval $I$, we extend
$\la$ to an endomorphism of $B(I)$.  This extension depends on
a choice of positive and negative crossings in the braiding
structure of representations of $\{A(I)\}$ and we denote it
with $\a_\la^\pm$ where $\pm$ stands for the choice of
positive and negative crossings.  This gives an ``almost''
representation of $\{B(I)\}$ and it is called a soliton
endomorphism.  This induction machinery is called $\a$-induction.
It was first introduced in \cite{LR1} and studied in detail in
\cite{X1}, \cite{BE}.  Ocneanu had a graphical calculus in a very different
context involving the $A$-$D$-$E$ Dynkin diagrams and the two
methods were unified in \cite{BEK1}, \cite{BEK2}.
It turns out that the intersection of irreducible endomorphisms
of $B(I)$ arising from $\a^+$-induction and $\a^-$-induction 
exactly gives those corresponding the representations of $\{B(I)\}$
by \cite{KLM}, \cite{BEK1}, \cite{BEK2}.

Let $\{A(I)\}$ be completely rational in the above setting.  Then
$\{B(I)\}$ is automatically also completely rational.  
(The converse also holds.)  The modular
tensor category of $\{A(I)\}$ gives a (finite dimensional) unitary
representation of $SL(2,{\mathbb{Z}})$ from its braiding.  (The
dimension of the representation is the number of irreducible
representations of $\{A(I)\}$ up to unitary equivalence.)
Define the matrix $(Z_{\la\mu})$ by 
$Z_{\la\mu}=\dim\Hom(\a^+_\la,\a^-_\mu)$ where $\la,\mu$ denote
endomorphisms of $A(I)$ corresponding to irreducible representations
of $\{A(I)\}$.  Then we have the following
in \cite{BEK1}.  

\begin{theorem}
The matrix $Z$ commutes with the above unitary
representation of $SL(2,{\mathbb{Z}})$.
\end{theorem}

Such $Z$ also satisfies $Z_{\la\mu}\in\{0,1,2,\dots\}$
and $Z_{00}=1$ where $0$ denotes the vacuum representation of
$\{A(I)\}$.  Such a matrix is called a modular invariant of the
representation of $SL(2,{\mathbb{Z}})$.  The number of modular
invariants for a given local conformal net $\{A(I)\}$ is always
finite and often quite limited.  This gives the following
classification method of all possible irreducible
extensions $\{B(I)\}$
for a given local conformal net $\{A(I)\}$.
(Any irreducible extension automatically has a finite index
by \cite{ILP}.)

\begin{enumerate}
\item Find all possible modular invariants $(Z_{\la\mu})$
for the modular tensor category arising from representations
of $\{A(I)\}$.
\item For each $(Z_{\la\mu})$, determine all possible $Q$-systems
corresponding to $\bigoplus Z_{0\la}\la$.
\item Pick up only local $Q$-systems.
\end{enumerate}

Consider a local conformal net $\{A(I)\}$.  The projective
unitary representation of $\Diff(S^1)$ gives a representation
of the Virasoro algebra and it gives a positive real number $c$ 
called the central charge.  This is a numerical invariant of
a local conformal net and the value of $c$ is in the set
$\{1-6/n(n+1)\mid n=3,4,5,\dots\}\cup [1,\infty)$.  We now
restrict ourselves to the case $c < 1$.  Let $\Vir_c(I)$ be
the von Neumann algebra generated by $U(g)$ where $g\in\Diff(S^1)$
acts trivially on $I'$.  This gives an extension 
$\{\Vir_c(I)\subset A(I)\}$.  It turns out $\{\Vir_c(I)\}$ is
completely rational and we can apply the above method to
classify all possible $\{A(I)\}$.  The modular invariants have
been classified in \cite{CIZ}, and locality and a certain
2-cohomology argument imply that the extensions exactly correspond
to so-called type I modular invariants.  We thus have a complete
classification of local conformal nets with $c < 1$ as follows
\cite{KL1}.

\begin{theorem}
Any local conformal net with $c < 1$ is one of the following.
\begin{enumerate}
\item The Virasoro nets $\{\Vir_c(I)\}$ with $c < 1$.
\item Their simple current extensions with index 2.
\item Four exceptionals at $c=21/22, 25/26, 144/145, 154/155$.
\end{enumerate}
\end{theorem}

The four exceptionals correspond to the Dynkin diagrams $E_6$ and $E_8$.
Three of them are identified with certain coset constructions, but
the remaining one with $c=144/145$ does not seem to be related to any
other known constructions so far.  All these four are given by an
extension by a $Q$-system.  Note that this appearance of modular
invariants is different from its original context in 2-dimensional
conformal field theory.

\section{Vertex operator algebras}

A vertex operator algebra gives another mathematical axiomatization of a 
chiral conformal field theory.  It deals with Fourier expansions
of operator-valued distributions, vertex operators, on $S^1$ in
an algebraic manner.

Recall that we have a complete list of 
finite simple groups today as follows \cite{FLM}.

\begin{enumerate}
\item Cyclic groups of prime order.
\item Alternating groups of degree $5$ or higher.
\item $16$ series of groups of Lie type over finite fields.
\item $26$ sporadic finite simple groups.
\end{enumerate}

The largest group among the 26 groups in the fourth in terms of the
order is called the Monster group, and its order is
approximately $8\times10^{53}$.
This group was first constructed by Griess.
It has been known that the smallest
dimension of a non-trivial irreducible representation
of the Monster group is 196883.

The next topic in this section is the
$j$-function.  This is a function of a complex
number $\tau$ with ${\mathrm{Im}}\;\tau>0$ with the
following expansion.
$$j(\tau)=q^{-1}+744+196884q+
21493760q^2+864299970 q^3+\cdots,$$
where we set $q=\exp(2\pi i \tau)$.

This function has modular invariance property
$$j(\tau)=j\left(\displaystyle\frac{a\tau+b}{c\tau+d}\right),$$
for
$$\left(\begin{array}{cc}
a & b \\ c& d\end{array}\right)\in SL(2,{\mathbb{Z}}),$$
and this property and the condition that the top term of the
Laurent series of $q$ start with $q^{-1}$ determine the $j$-function
uniquely except for the constant term.

McKay noticed that the first non-trivial coefficient of
the Laurent expansion of the $j$-function except for the constant
term is 196884 which is ``almost'' 196883.
Extending this idea, Conway-Norton \cite{CN} formulated 
the Moonshine conjecture as follows.

\begin{conjecture}
\begin{enumerate}
\item
We have some graded infinite dimensional $\mathbb C$-vector space
$V=\bigoplus_{n=0}^\infty V_n$ $(\dim V_n<\infty)$
with some natural algebraic structure and its automorphism
group is the Monster group.
\item Each element $g$ of the Monster group acts
on each $V_n$ linearly.  The Laurent series
$$\sum_{n=0}^\infty ({\mathrm{Tr}}\;g|_{V_n})q^{n-1}$$
arising from the trace value of the $g$-action on $V_n$
is a classical function called a Hauptmodul corresponding
to a genus 0 subgroup of $SL(2,{\mathbb{R}})$.
(The case $g$ is the identity element is the $j$-function
without the constant term.)
\end{enumerate}
\end{conjecture}

``Some natural algebraic structure'' in the above conjecture has
been formulated as a vertex operator algebra in \cite{FLM}
and the full Moonshine conjecture has been proved by
Borcherds \cite{B}.
The axioms of a vertex operator algebra are given as follows.

Let $V$ be a $\mathbb C$-vector space. We say that a formal
series $a(z) =
\sum_{n\in\Z} a_{(n)} z^{-n-1}$ with coefficients
$a_{(n)} \in \End(V)$ is a field on $V$, 
if for any $b \in V$,  we have $a_{(n)}b = 0$ for all
sufficiently large $n$. 

A $\mathbb C$-vector space $V$ is called a vertex algebra
if we have the following properties.
\begin{enumerate}
\item (State-field correspondence) For each $a\in V$, we have a
field $Y(a,z)=\sum_{n\in\Z} a_{(n)} z^{-n-1}$ on $V$.
\item (Translation covariance)
We have a linear map $T\in \End(V)$ such that we have
$[T,Y(a,z)]=\frac{d}{dz}Y(a,z)$ for all $a\in V$.
\item (Existence of the vacuum vector)
We have a vector $\Omega\in V$ with
$T\Omega=0$, $Y(\Omega,z)=\id_V$, $a_{(-1)}\Omega=a$.
\item (Locality) For all $a,b\in V$, we have
$(z-w)^N[Y(a,z),Y(b,w)]=0$ for a sufficiently large
integer $N$.
\end{enumerate}
We then call $Y(a,z)$ a vertex operator.
(The locality axiom
is one representation of the idea that
$Y(a,z)$ and $Y(b,w)$ should commute for $z\neq w$.)

Let $V$ be a $\mathbb C$-vector space and $L(z)=\sum_{n\in\Z}L_n z^{-n-2}$ be
a field on $V$.  If the endomorphisms $L_n$ satisfy the Virasoro algebra
relations
$$[L_m,L_n]=(m-n)L_{m+n}+\frac{(m^3-m)\delta_{m+n,0}}{12}c,$$
with central charge $c\in\mathbb C$, then we say $L(z)$ is a Virasoro
field.  If $V$ is a vertex algebra and
$Y(\omega,z)=\sum_{n\in\Z} L_n z^{-n-2}$ is a Virasoro
field, then we say $\omega\in V$ is a  Virasoro vector.
A Virasoro vector $\omega$ is called a conformal vector if
$L_{-1}=T$ and $L_0$ is diagonalizable on $V$, that is,
$V$ is an algebraic direct sum of the eigenspaces of $L_0$.
Then the corresponding vertex operator $Y(\omega,z)$ is called the
energy-momentum field and $L_0$ the conformal Hamiltonian.
A vertex algebra with a conformal vector is called a 
conformal vertex algebra.
We then say $V$ has central charge $c\in\C$.

A nonzero element $a$ of a conformal vertex algebra in $\Ker(L_0-\a)$ is
said to be a homogeneous element of conformal weight
$d_a=\a$.  We then set $a_n=a_{(n+d_a-1)}$ for $n\in{\mathbb{Z}}-d_a$.
For a sum $a$ of homogeneous elements, we extend $a_n$ by linearity.

A homogeneous element $a$ in a conformal vertex algebra $V$ and the
corresponding field $Y(a,z)$ are called quasi-primary if $L_1 a=0$
and primary if $L_na=0$ for all $n>0$.

We say that a conformal vertex algebra $V$ is of CFT type
if we have $\Ker(L_0-\a)\neq0$ only for $\a\in\{0,1,2,3,\dots\}$
and $V_0=\C\Omega$.

We say that a conformal vertex algebra $V$ is a
vertex operator algebra if we have the following.
\begin{enumerate}
\item We have $V=\bigoplus_{n\in\Z} V_n$, where $V_n=\Ker(L_0-n)$.
\item We have $V_n=0$ for all sufficiently small $n$.
\item We have $\dim(V_n)<\infty$ for $n\in\mathbb Z$.
\end{enumerate}

Basic sources of constructing vertex operator algebras
are affine Kac-Moody and Virasoro algebras due to Frenkel-Zhu
and even lattices due to Frenkel-Lepowsky-Meurman.
Methods to construct new examples from known
examples are a tensor product, a simple current extension
due Schellekens-Yankielowicz,
orbifold construction due to Dijkgraaf-Vafa-Verlinde-Verlinde,
coset construction due to Frenkel-Zhu, and an extension
by a Q-system due to Huang-Kirillov-Lepowsky.
These are parallel to constructions of
local conformal nets, but constructions of vertex operator
algebras are earlier except for the extension by a Q-system.

A representation theory of a vertex operator algebra is
known as a theory of modules.  It has been shown by Huang
that we have a modular tensor category for a well-behaved
vertex operator algebra.  (The well-behavedness condition
is basically the so-called $C_2$-cofiniteness.)

\section{From a vertex operator algebra to a local
conformal net and back}

We now would like to construct a local conformal net from a vertex
operator algebra $V$.  First of all, we need a Hilbert space
of states, and it should be the completion of $V$ with respect to
some natural inner product.  A vertex operator algebra with such
an inner product is called unitary.  Many vertex operator algebras
are unitary, but also many others are non-unitary.  In order to
have the corresponding local conformal net, we definitely have
to assume that $V$ is unitary.  We now give a precise definition
of a unitary vertex operator algebra.

An invariant bilinear form on a vertex operator algebra $V$ is a
bilinear form $(\cdot,\cdot)$ on $V$ satisfying
$$(Y(a,z)b,c)=(b,Y(e^{zL_1}(-z^{-2})^{L_0}a,z^{-1})c)$$
for all $a,b,c\in V$.

For a vertex operator algebra $V$ with a conformal vector $\omega$,
an automorphism $g$ as a vertex algebra is called a VOA 
automorphism if we have $g(\omega)=\omega$.

Let $V$ be a vertex operator algebra and suppose we have a
positive definite inner product $(\cdot\mid\cdot)$, where we
assume this is antilinear in the first variable.  We
say the inner product is normalized if we have $(\Omega\mid\Omega)=1$.
We say that the inner product is invariant if there
exists a VOA antilinear automorphism $\theta$ of $V$ such that
$(\theta\cdot\mid\cdot)$ is an invariant bilinear form on $V$.
We say that $\theta$ is a PCT operator associated with
the inner product.

If we have an invariant inner product, we automatically have
$(L_n a\mid b)=(a\mid L_{-n}b)$ for $a,b\in V$ and also
$V_n=0$ for $n<0$.  The PCT operator $\theta$ is unique and
we have $\theta^2=1$ and $(\theta a\mid \theta b)=(b\mid a)$
for all $a,b\in V$. (See \cite[Section 5.1]{CKLW} for details.)

A unitary vertex operator algebra $V$ is a pair of
a vertex operator algebra and a normalized invariant
inner product. It is simple if we have $V_0=\C\Omega$.

Now suppose $V$ is a unitary vertex operator algebra.  A
vertex operator $Y(a,z)$ should mean a Fourier expansion of
an operator-valued distribution on $S^1$.  For a test function
$f$ with Fourier coefficients $\hat f_n$, the action of the
distribution $Y(a,z)$ applied to the test function $f$ on
$b\in V$ should be $\sum_{n\in\Z} \hat f_n a_n b$.  In order
to make sense out of this, we need convergence of this 
infinite sum.  To insure such convergence, we introduce 
the following notion of energy-bounds.

Let $(V, (\cdot\mid\cdot))$ be a unitary vertex operator
algebra.  We say that $a\in V$ (or $Y(a,z)$) satisfies 
energy-bounds if we have positive integers $s,k$
and a constant $M>0$ such that we have
$$\|a_nb\|\le M(|n|+1)^s\|(L_0+1)^kb\|,$$
for all $b\in V$ and $n\in\mathbb Z$.
If every $a\in V$ satisfies energy-bounds, we say
$V$ is energy-bounded.

We have the following Proposition in \cite{CKLW}.

\begin{proposition}
If $V$ is a simple unitary 
vertex operator algebra generated by $V_1$ and $F\subset V_2$
where $F$ is a family of quasi-primary $\theta$-invariant 
Virasoro vectors, then $V$ is energy-bounded.
\end{proposition}

We now assume $V$ is energy-bounded.  Let $H$ be the completion
of $V$ with respect to the inner product.  For any $a\in V$ and
$n\in\mathbb Z$, we regard $a_{(n)}$ as a densely defined operator on
$H$.  This turns out to be closable.
Let $f(z)$ be a smooth 
function on $S^1=\{z\in\C\mid |z|=1\}$ with Fourier coefficients
$$\hat f_n=\int_{-\pi}^\pi f(e^{i\theta})e^{-in\theta}
\frac{d\theta}{2\pi}$$
for $n\in\Z$.  For every $a\in V$, we define the operator
$Y_0(a,f)$ with domain $V$ by
$$Y_0(a,f)b=\sum_{n\in\Z} \hat f_n a_n b$$
for $b\in V$.  The convergence follows from the energy-bounds
and $Y_0(a,f)$ is a densely defined operator.  This is again
closable.  We denote by $Y(a,f)$ the closure of $Y_0(a, f)$
and call it a smeared vertex operator.

We define  $\A_{(V,(\cdot\mid\cdot))}(I)$ 
to be the von Neumann algebra generated by the
(possibly unbounded) operators $Y(a,f)$ with
$a\in V$, $f\in C^\infty(S^1)$ and $\supp\;f\subset I$.
The family $\{\A_{(V,(\cdot\mid\cdot))}(I)\}$ clearly
satisfies isotony.  We can verify that 
$(\bigvee_I \A_{(V,(\cdot\mid\cdot))}(I))\Omega$
is dense in $H$.  A proof 
of conformal covariance is nontrivial, but can be done
as in \cite{TL} by studying the representations
of the Virasoro algebra and $\Diff(S^1)$.  We also have the vacuum
vector $\Omega$ and the positive energy condition.
However, locality is not clear at all from our construction, 
so we make the following definition.

We say that a unitary vertex operator algebra
$(V,(\cdot\mid\cdot))$ is strongly local if
it is energy-bounded and we have
$\A_{(V,(\cdot\mid\cdot))}(I)\subset 
\A_{(V,(\cdot\mid\cdot))}(I')'$ for all intervals 
$I\subset S^1$.

A strongly local unitary vertex operator algebra
produces a local conformal net through the above
procedure by definition, but the definition of strong
locality looks like we assume what we want to prove, 
and it would be useless unless we have a good criterion
for strong locality.  The following theorem gives such
a criterion \cite{CKLW}.

\begin{theorem}
Let $V$ be a simple unitary vertex operator algebra
generated by $V_1 \cup F$ where $F\subset V_2$ is a family of
quasi-primary $\theta$-invariant Virasoro vectors, then
$V$ is strongly local.
\end{theorem}

The above criteria applies to vertex operator algebras arising from
the affine Kac-Moody and Virasoro algebras.
We also have the following theorem which we can apply to
many examples \cite{CKLW}.

\begin{theorem}
(1) Let $V_1, V_2$ be simple unitary strongly local vertex operator algebras.
Then $V_1\otimes V_2$ is also strongly local.

(2) Let $V$ be a simple unitary strongly local vertex operator algebra
and $W$ its subalgebra.  Then $W$ is also strongly local.
\end{theorem}

The second statement of the above theorem shows that strong locality
passes to orbifold and coset constructions, in particular.

For a unitary vertex operator algebra $V$,
we write $\Aut(V)$ for the automorphism
group of $V$.  For a local conformal net $\{A(I)\}$, we have
a notion of the automorphism group and we write
$\Aut(A)$ for this.  We have the following in \cite{CKLW}.

\begin{theorem}
Let $V$ be a strongly local unitary vertex operator algebra
and $\{A_{(V,(\cdot\mid\cdot))}(I)\}$ the corresponding
local conformal net.  Suppose $\Aut(V)$ is finite.  
Then we have
$\Aut(A_{(V,(\cdot\mid\cdot))})=\Aut(V)$.
\end{theorem}

The Moonshine vertex operator algebra $V^\natural$ is
strongly local and unitary, so we can apply the above result
to this to obtain the Moonshine net.  It was first constructed in
\cite{KL4} with a more ad-hoc method.

For the converse direction, we have the following \cite{CKLW}.

\begin{theorem}
Let $V$ be a simple unitary strongly local 
vertex operator algebra and $\{\A_{(V,(\cdot\mid\cdot))}(I)\}$
be the corresponding local conformal net.
Then one can recover the vertex operator algebra
structure on $V$, which is an algebraic direct sum of
the eigenspaces of the conformal Hamiltonian,
from the local conformal net
$\{\A_{(V,(\cdot\mid\cdot))}(I)\}$.
\end{theorem}

This is proved by using the Tomita-Takesaki theory and
extending the methods in \cite{FJ}.  Establishing
correspondence between the representation theories of
a vertex operator algebra and a local conformal net is
more difficult, though we have some recent progress due
to Carpi, Weiner and Xu.  The method of \cite{T} may be more
useful for this.  We list the following conjecture
on this.  (For a representation of a local conformal net,
we define the character as $\Tr(q^{L_0-c/24})$ when it
converges for some small values of $q$.  We have a similar
definition for a module of a vertex operator algebra.)

\begin{conjecture}\label{rep}
We have a bijective correspondence between 
completely rational local conformal nets and
simple unitary $C_2$-cofinite vertex operator algebras.
We also have equivalence of tensor categories for
finite dimensional representations of a 
completely rational local conformal net and
modules of the corresponding vertex operator
algebra.  We further have coincidence of
the corresponding characters of the
irreducible representations
of a completely rational
local conformal net and irreducible
modules of the corresponding vertex operator algebra.
\end{conjecture}

Recall that we have a classical correspondence between Lie algebras and Lie
groups.  The correspondence between affine Kac-Moody algebras
and loop groups is similar to this, but ``one step higher''.
Our correspondence between vertex operator algebras and
local conformal nets is something even one more step higher.

Finally we discuss the meaning of strong locality.  We have no 
example of a unitary vertex operator algebra which is known to
be not strongly local.  If there should exist such an example,
it would not correspond to a chiral conformal field theory in
a physical sense.  This means that one of the following holds:
any simple unitary vertex operator algebra is strongly local or 
the axioms of unitary vertex operator algebras are too weak
to exclude non-physical examples.

\section{Other types of conformal field theories}

Here we list operator algebraic treatments of conformal field theories
other than chiral ones.

Full conformal field theory is a theory on the 2-dimensional
Minkowski space.  We axiomatize a net of von Neumann algebras
$\{B(I\times J)\}$ 
parameterized by double cones (rectangles) in the Minkowski
space in a similar way to the case of local conformal nets.
From this, a restriction procedure produces two local conformal
nets $\{A_L(I)\}$ and $\{A_R(I)\}$.  We assume both are completely
rational.  Then we have a subfactor
$A_L(I)\otimes A_R(J)\subset B(I\times J)$ which automatically
has a finite index, and the study of $\{B(I\times J)\}$ is
reduced to studies of $\{A_L(I)\}$, $\{A_R(I)\}$ and this
subfactor.  A modular invariant again naturally appears here
and we have a general classification theory.  For the case
of central charge less than $1$, we obtain a complete and
concrete classification result as in \cite{KL2}.

A boundary conformal field theory is a quantum field theory
on the half-Minkowski space $\{(t,x)\in\M \mid x>0\}$.
The first general theory to deal with this setting was
given in \cite{LR2}.  We have more results in
\cite{CKL2} and \cite{BKL}.  In this case, a restriction
procedure gives one local conformal net.  We assume that
this is completely rational.  Then we have a non-local,
but relatively local
extension of this completely rational local conformal net
which automatically has a finite index.  The study of
a boundary conformal field theory is reduced to studies
of this local conformal net and a non-local extension.
For the case
of central charge less than $1$, we obtain a complete and
concrete classification result as in \cite{KLPR} along the
line of this general theory.

We also have results on the
phase boundaries and topological defects
in the operator algebraic
framework in  \cite{BKLR1}, \cite{BKLR2}.
See \cite{FuRS} for earlier works on topological defects.

A superconformal field theory is a version of 
${\mathbb {Z}}_2$-graded conformal field theory having extra
supersymmetry.
We have operator algebraic versions of $N=1$ and $N=2$
superconformal field theories as in 
\cite{CKL1} and \cite{CHKLX} based on $N=1$ and $N=2$
super Virasoro algebras, and there we 
have superconformal nets rather than local conformal
nets.  We also have relations of this theory to
noncommutative geometry in \cite{KL3},
\cite{CHKL}, \cite{CHKLX}.

\section{Future directions}

We list some problems and conjectures for the future studies
at the end of this article.

\begin{conjecture}
For a completely rational local conformal net, we have
convergent characters for all irreducible representations
and they are closed under modular transformations of $SL(2,\Z)$.
Furthermore, the $S$-matrix defined with braiding gives
transformation rules of the characters under the transformation
$\tau\mapsto -1/\tau$.
\end{conjecture}

This conjecture was made in \cite[page 625]{FG} and follows
from Conjecture \ref{rep}.  

We say that a local conformal net is holomorphic if its
only irreducible representation is the vacuum representation.
The following is \cite[Conjecture 3.4]{X5} which is the operator
algebraic counterpart of the famous uniqueness conjecture of
the Moonshine vertex operator algebra.

\begin{conjecture}
A holomorphic local conformal net 
with $c=24$ and the eigenspace of
$L_0$ with eigenvalue $1$ being $0$ is unique up to 
isomorphism.
\end{conjecture}

A reason to expect such uniqueness from an operator algebraic
viewpoint is that a set of  simple algebraic invariants should
be a complete invariant as long as we have some kind of amenability,
which is automatic in the above case.

The following is an operator algebraic counterpart of
\cite[Conjecture 3.5]{Ho}.

\begin{conjecture}
Fix a modular tensor category $\mathcal C$ and a central charge $c$.
Then we have only finitely many local conformal nets with representation
category $\mathcal C$ and central charge $c$.
\end{conjecture}

From an operator algebraic viewpoint, the following problem is also
natural.

\begin{problem}
Suppose a finite group $G$ is given.  Construct a local conformal
net whose automorphism group is $G$ in some canonical way.
\end{problem}

This ``canonical'' method should produce the Moonshine net if
$G$ is the Monster group.  We may have to consider some
superconformal nets rather than local conformal nets to get a
nice solution.

Conformal field theory on Riemann surfaces 
has been widely studied and conformal blocks play a important
role there.  It is not clear
at all how to formulate this in our operator algebraic
approach to conformal field theory, so we have the
following problem.

\begin{problem}
Formulate a conformal field theory on a Riemann surface in the
operator algebraic approach.
\end{problem}

It is expected that the $N=2$ full superconformal field
theory is related to Calabi-Yau manifolds, so
we also list the following problem.

\begin{problem}
Construct an operator algebraic object corresponding
to a Calabi-Yau manifold in the setting of 
$N=2$ full superconformal field theory
and study the mirror symmetry in this context.
\end{problem}

The structure of a modular tensor category naturally
appears also in the context of
topological phases of matters and anyon condensation
as in \cite{K2}, \cite{K3}, \cite{Ko}.
(The results in \cite{BEK3} can be also seen in this
context.)  We list the following problem.

\begin{problem}
Relate local conformal nets directly with 
topological phases of matters and anyon condensation.
\end{problem}

\end{document}